\documentclass[a4paper]{ESASPCS13Style}
\usepackage{epsfig}
\newcommand{\teff}{T_{\rm eff}}
\newcommand{\kms}{km\,s$^{-1}$}
\newcommand{\LiI}{Li\,{\sc i}}
\newcommand{\FeI}{Fe\,{\sc i}}
\begin{document}

\title{Late-type field stars in the RASS at high galactic latitude}

\author{F.-J. Zickgraf\inst{1} \and J. Krautter \inst{2} \and  
S. Frink \inst{3} \and J.M. Alcal\'a \inst{4} \and R.
Mujica \inst{5}  \and E. Covino,\inst{4}  \and M.F. Sterzik \inst{6}} 
\institute{Hamburger Sternwarte, Gojenbergsweg 112, 21029 Hamburg, Germany
\and Landessternwarte K\"onigstuhl, D-69117 Heidelberg, Germany
\and Sterrewacht Leiden, P.O. Box 9513, NL-2300 RA Leiden, The Netherlands
\and Osservatorio Astronomico di Capodimonte, Via Moiariello 16, I-80131 Napoli,
Italy
\and Instituto Nacional de Astrofisica, Optica y Electronica, 
A. Postal 51 y 216 Z.P., 72000 Puebla, Mexico
\and European Southern Observatory, Alonso de Cordova 3107, Santiago 19, Chile}

\maketitle 

\begin{abstract}

We present results of an investigation on a high-galactic 
latitude sample of late-type field stars selected from 
the ROSAT All-Sky Survey (RASS). The sample comprises 
$\sim$200 G, K, and M stars. Lithium abundances were determined 
for  $\sim$180 G-M stars. Radial velocities were measured for 
most of the $\sim$140 G and K type stars. 
Combined with proper motions these data were used to study the age 
distribution and the kinematical properties of the sample. 
Based on the lithium abundances half of the G-K stars were 
found to be younger than the Hyades (660\,Myr). About 30\% are 
comparable in age to the Pleiades (100 \,Myr). A small 
subsample of 10 stars is younger than the Pleiades (100\,Myr). 
They are therefore most likely pre-main sequence stars. 
Kinematically the PMS and Pleiades-type stars appear to form a group 
with space velocities close to the Castor moving group but 
clearly distinct from the Local Association. 

\end{abstract}

\section{Introduction}
\begin{figure}[ht]
  \begin{center}
    \epsfig{file=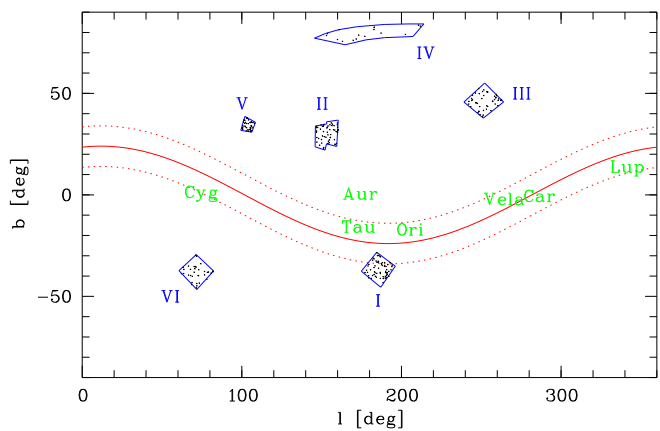,angle=-90,width=8.7cm}
  \end{center}
\caption{Location of the 6 study areas in galactic coordinates. Also shown is the Gould
Belt according to Guillout et al. (1998). }
\label{area}
\end{figure}

In a previous investigation a sample of RASS X-ray sources located 
at $|b| \ge 20^{\circ}$ and $DEC \ge -9^{\circ}$ 
has been identified optically (\cite{Zi97}). 
The sample comprises 674 X-ray sources distributed in 6 study areas 
with a total area of 685 deg$^{2}$ (see Fig. \ref{area}). 
The sample contains 273 stellar  X-ray emitters  
of spectral type F to M plus 20 other stellar sources (CVs, WDs, A stars).
The rest are  extragalactic X-ray emitters and a few unidentified sources.
The catalogue of identifications
and the statistical analysis  can be found in \cite*{App98} and \cite*{Kraut99},
respectively. Here we present the results of an investigation of the X-ray 
properties, the age distribution and the kinematics 
of the stellar subsample of coronal emitters.  

\section{Observations}
\label{sec:obs}
The stellar counterparts of spectral types G to M were observed spectroscopically
in the red wavelength region containing the line \LiI\,$\lambda$6708\AA\ during several
observing campaigns between 1996 and 2002 mainly at Calar Alto Observatory. 
Further observations were obtained at Obs. Haute Provence and at ESO, La Silla.
F stars were omitted because the \LiI\ line is not a sensitive  age  indicator
for these stars. We observed 172 of 199  stars with spectral types G to M. 
Data for 6 further G-K stars were found in the literature. 
Of the 141 G and K stars 118 were sufficiently bright ($V \le $ 12$^{\mbox m}$) 
for high-resolution spectroscopy with  a resolution of
$\Delta \lambda \sim$ 0.2 - 0.3\,\AA\  at 6708\,\AA . Likewise, 7 M stars could be
observed with high resolution. Fainter stars were observed with low spectral 
resolution of $\Delta \lambda \sim$ 3-4\,\AA .
Seven 7 G-K stars and 13 M stars are still unobserved.
In total, spectroscopic observations of 179 out of 199 G, K, and M stars are available.

Radial  velocities (RVs) and projected rotational velocities, $v\,\sin i$, 
were derived from the high-resolution spectra by means of cross-correlation 
techniques.  

\section{Results}
\label{sec:results}
\subsection{Distances}
The  distances towards the G and K stars were derived from trigonometric or 
spectroscopic parallaxes. 
The high-resolution spectra allowed the determination of the luminosity classes 
and hence of the absolute  visual magnitudes, $M_V$.
For the classification procedure spectra of MK standard stars were taken 
from the stellar library of \cite*{Pru01}. 
In this way spectroscopic parallaxes could be derived for 75 G-K stars.
For M stars we obtained IR photometric parallaxes.  43 stars
have IR photometry in the 2MASS catalogue. In the (J-H)-(H-K) diagramme
the  M stars are located along the track for main-sequence stars. 
Distances were therefore derived for $M_V$ magnitudes of main-sequence stars.
Distances for 26 G-K stars and 4 M stars were found in the Hipparcos catalogue.
For further 7 M stars trigonometric parallaxes were found in \cite*{GlieseJahreiss91}.
In total, distances could be determined for 101 G-K and 54 M stars.
For the remaining stars we assumed luminosity class V for the estimation of a lower
limit for the distance.

The distribution of distances shows a maximum around 50\,pc and a tail extending 
up to several 100\,pc with most stars being nearer than 300\,pc. 

\subsection{X-ray luminosities}
\begin{figure}[ht]
  \begin{center}
    \epsfig{file=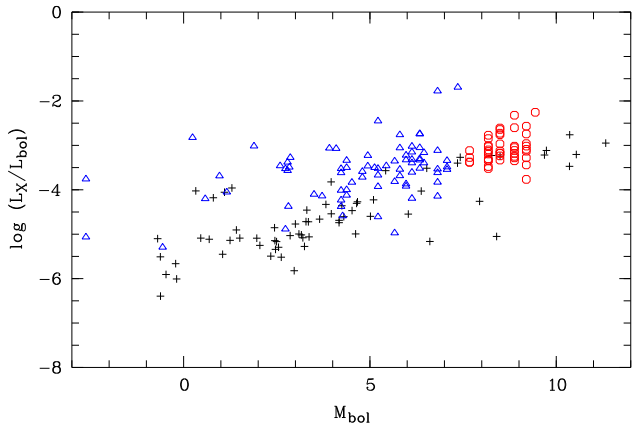,angle=-90,width=8.7cm}
  \end{center}
\caption{Ratio of X-ray and bolometric luminosity for all single stars with
trigonometric ($+$ sign), spectroscopic (triangles), or IR photometric parallaxes 
(circles) as a function of bolometric magnitude, $M_{\rm bol}$. }
\label{lxlbol}
\end{figure}

Using these distances  X-ray luminosities $L_X$(0.1-2.4\,keV) and bolometric 
luminosities $L_{\rm bol}$ were calculated.  The ratio $L_X/L_{\rm bol}$ is shown in 
Fig. \ref{lxlbol} as a function of bolometric magnitude, $M_{\rm bol}$. The highest
ratio is measured for the latest spectral types which have the faintest $M_{\rm bol}$.

\subsection{Lithium abundances}
\begin{figure}[ht]
  \begin{center}
    \epsfig{file=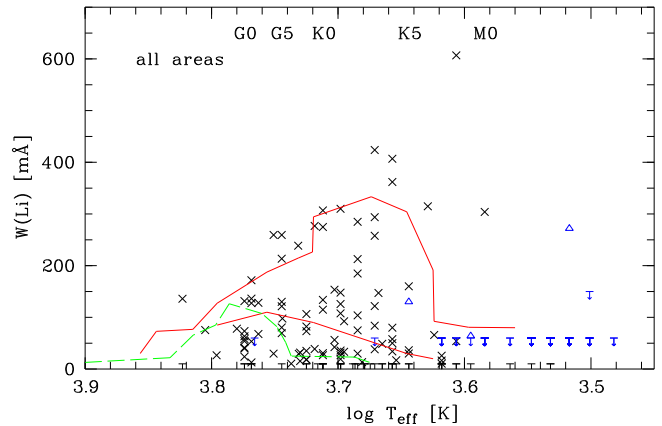,angle=-90,width=8.7cm}
  \end{center}
\caption{Equivalent widths of \LiI\,$\lambda6708$ as a function of $\teff$ 
for all stars in the six study areas. Crosses and triangles are high and 
low resolution measurements, respectively. 
The solid lines denote the upper and lower envelope of the
lithium equivalent widths in the Pleiades adopted from Soderblom et al. (1993). 
The dashed line shows the upper  envelope for the Hyades cluster 
taken from  Thorburn et al. (1993).
}
\label{wli}
\end{figure}

\begin{figure}[ht]
  \begin{center}
    \epsfig{file=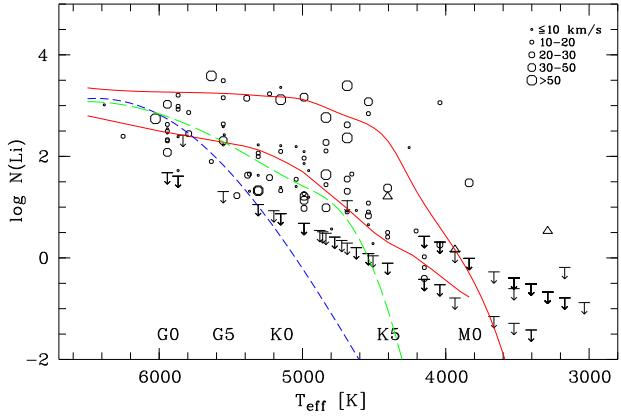,angle=-90,width=8.7cm}
  \end{center}
\caption{Lithium abundances versus effective temperature for the complete sample. 
Upper limits are plotted as
downward arrows. Circles  denote high-resolution measurements with the symbol size
depending on $v\,\sin i$. Low and medium resolution data are plotted as 
triangles. The solid lines are the upper and lower limit of $\log N$(\LiI) in the
Pleiades; the long dashed and short dashed lines show the upper $\log N$(\LiI)
limits for the UMaG and the Hyades, respectively.  
}
\label{nli}
\end{figure}

Equivalent widths of \LiI\,$\lambda$6708, $W$(Li), were measured in high 
resolution spectra by direct integration. The contribution of 
\FeI\,$\lambda$6708.44\AA\ was
taken into account. The detection limit was $\sim$10\,m\AA .
In low/medium resolution spectra \LiI\,$\lambda$6708 is blended with 
\FeI\,$\lambda\lambda$6703, 6705, and 6710\,\AA . The contributions of the 
individual lines were separated by means of the fitting method described by 
Zickgraf et al. (1998).
The detection limit in the low/medium resolution spectra is 
$\sim$60\,m\AA\ for K stars and $\sim$200\,m\AA\ for M stars. 
The equivalent widths are displayed in Fig. \ref{wli} as function of effective
temperature, $\teff$. 
Conversion of $W$(Li) to abundances $N$(Li) was performed using the curves of growth
of \cite*{Soder93} for $\teff \ge 4000$\,K and \cite*{Pav96} for cooler stars. 
We applied the spectral type - $\teff$ relations of 
de Jager \& Nieuwenhuijzen (1987).
The estimated temperature uncertainty is $\sim$200\,K which leads to an average
error of $N$(Li) of $\approx$ 0.3\,dex. In Fig. \ref{nli} $N$(Li) 
is displayed  as function of $\teff$. Included in the figure are the 
upper and lower envelope of $N$(Li) for the Pleiades, upper envelopes for
the Ursa Major Moving Group (UMaG) and for the  Hyades, respectively.

\subsection{Age estimates}
Age groups were defined based on lithium abundances,  $N$(Li), and on the 
comparison with the $N$(Li) distribution in the  Pleiades (100\,Myr), UMaG 
(300\,Myr) and Hyades (660\,Myr).
The age group PMS contains stars with $N$(Li) a\-bove the Pleiades upper envelope,
and the age group Pl\_ZAMS comprises stars with $N$(Li) between the Pleiades upper and 
lower envelope. 
The age group UMa consists of stars  between the Pleiades lower envelope and Hy\-ades 
upper envelope, and  
age group Hya+ has $N$(Li) below the Hy\-ades upper envelope.
The assigned ages and fractions of G-K stars in each age group are summarized in 
Table \ref{age} together with the median rotational velocity, $v\sin i$ of the
respective age group. Obviously,  $v\,\sin i$ decreases with increasing age.

\begin{table} 
\tabcolsep=4pt
\caption[]{Statistics of the age distribution of the sample of G-K stars.  
The total number of G-K stars is 141. The third line gives the median  
rotational velocity in \kms .
} 
\begin{tabular}{lllll} 
\noalign{\smallskip}     
\hline 
\noalign{\smallskip}    
       & \multicolumn{4}{c}{age group} \\ 
\noalign{\smallskip}     
       & PMS & Pl\_ZAMS & UMa  &  Hya+ \\ 
       & $<100$\,Myr & 100\,Myr & $\sim300$\,Myr & $>660$\,Myr \\ 
\hline 
\noalign{\smallskip}    
number & 8 & 40  & 19  & 74 \\ 
fraction& 6\% & 28\%  & 13\%  & 52\% \\ 
${\langle v\sin i \rangle}_{\rm med}$   &  32  &  17 & 18 & 11\\ 
\noalign{\smallskip}    
\hline 
\noalign{\smallskip}    
\end{tabular}
\label{age}
\end{table}

\subsection{Age dependent $\log N - \log S$ distribution}
\begin{figure}[ht]
  \begin{center}
    \epsfig{file=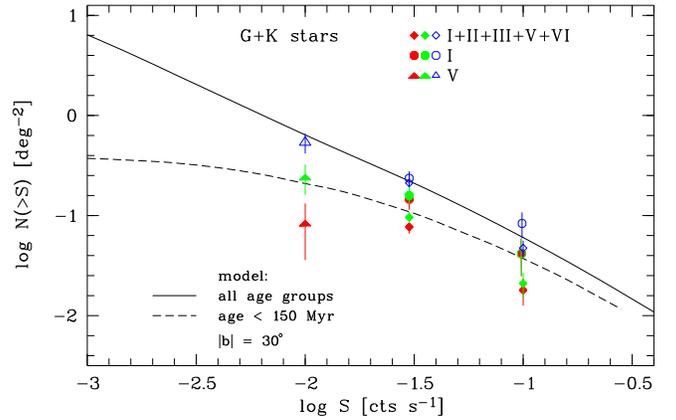,angle=-90,width=8.7cm}
  \end{center}
\caption{Comparison of observed coadded number densities of G and K stars, 
$N(>S)$, for three RASS count 
rates $S$ with models of Guillout et al. (1996) for $|b| = 30^{\circ}$. 
Open symbols denote the sum of all age groups. Lower and upper filled symbols 
represent  the sum of age groups PMS and
Pl\_ZAMS, and of PMS, Pl\_ZAMS, and UMa, respectively. 
}
\label{guillout}
\end{figure}

We  compared the observed cumulative number distribution, 
$\log N(>S) - \log S$, of our sample with model predictions by \cite*{Guillout96}. 
The models give cumulative surface densities, $N(>S)$, as a function
of ROSAT-PSPC count rate, $S$, for three age bins:  age younger
then 150\,Myr, age between 150\,Myr and 1\,Gyr, and older than 1\,Gyr. We  restricted
the comparison to the youngest model age bin and to the sum
of all model age bins because of the difficulty to separate observationally 
stars with ages of several 100\,Myr to $\sim1$\,Gyr and older.
In Fig. \ref{guillout} the comparison is shown for the stars in the five study 
areas located around $|b| = 30^{\circ}$. Shown are the sum of all age groups
and of the sum of age groups PMS, PL\_ZAMS, and of PMS, PL\_ZAMS, and UMa 
which represent the subsample younger than about 150\,Myr.
The comparison shows that the observed numbers of stars are in
reasonably good agreement with the model predictions. 
This holds for both the sum of all age groups and stars younger than
$\sim150$\,Myr obtained as described above and represented in the figure 
by the filled symbols. Likewise, the predicted flattening of $\log N(>S) - \log S$ 
at increasinging smaller count rates is also found in our data for area V 
which has the lowest count rate limit of 0.01\,cts\,s$^{-1}$. 

\subsection{Kinematics}
Proper motions were obtained  for 138 G, K, and M stars from the Hipparcos, 
TYCHO-2, UCAC2, and STARTNET catalogues.
Space velocity components $U$, $V$, and $W$ were calculated 
in the Local Standard of Rest (LSR) (right-handed coordinate system)
for 89 single G-K and 8 M stars  with distance, RV, and proper motion. 
For the correction of the solar motion we used the solar motion vector of 
\cite*{Dehnen98}. 

\begin{figure}[ht]
  \begin{center}
    \epsfig{file=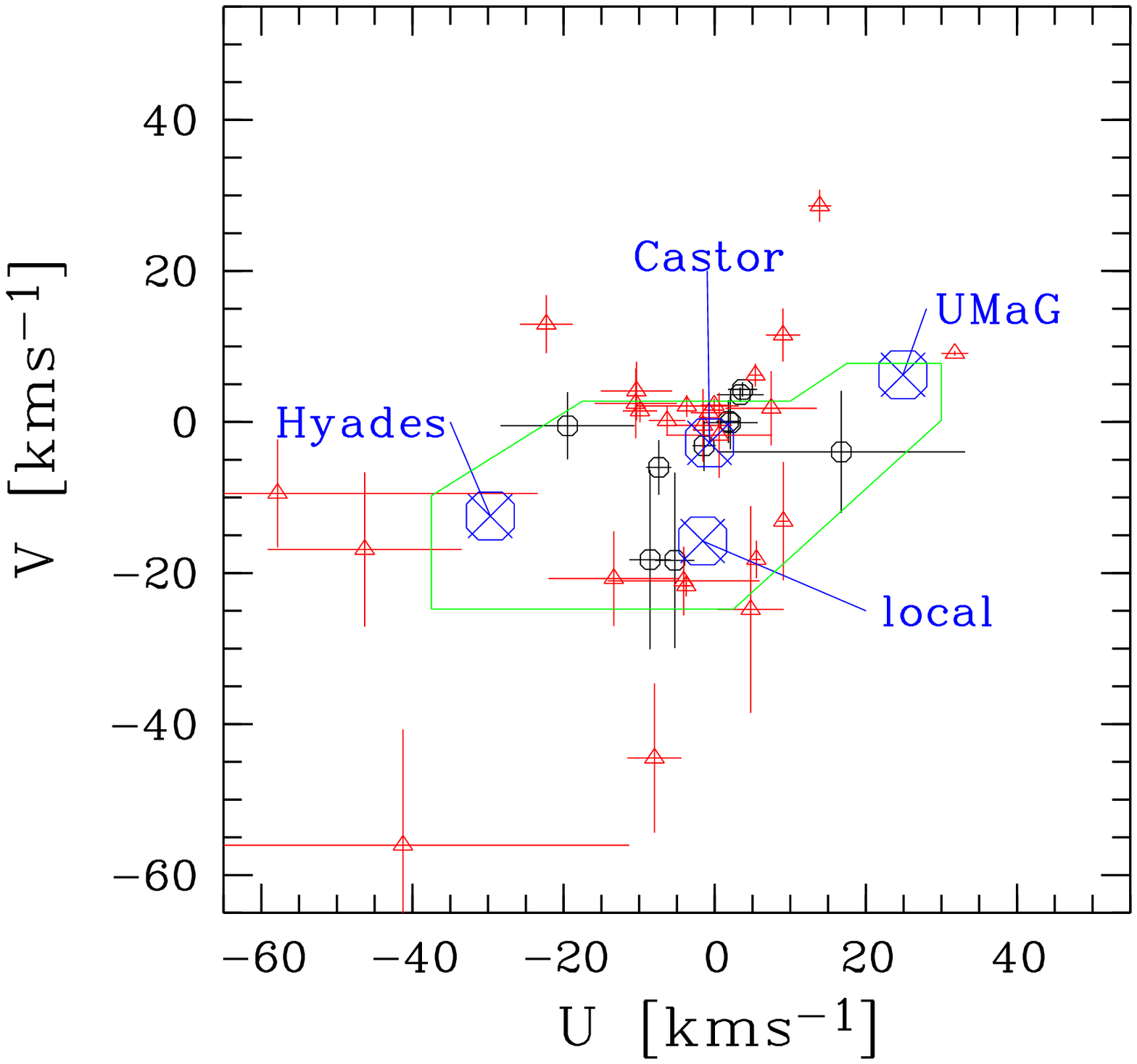,angle=-90,width=8.7cm}
  \end{center}
  \begin{center}
    \epsfig{file=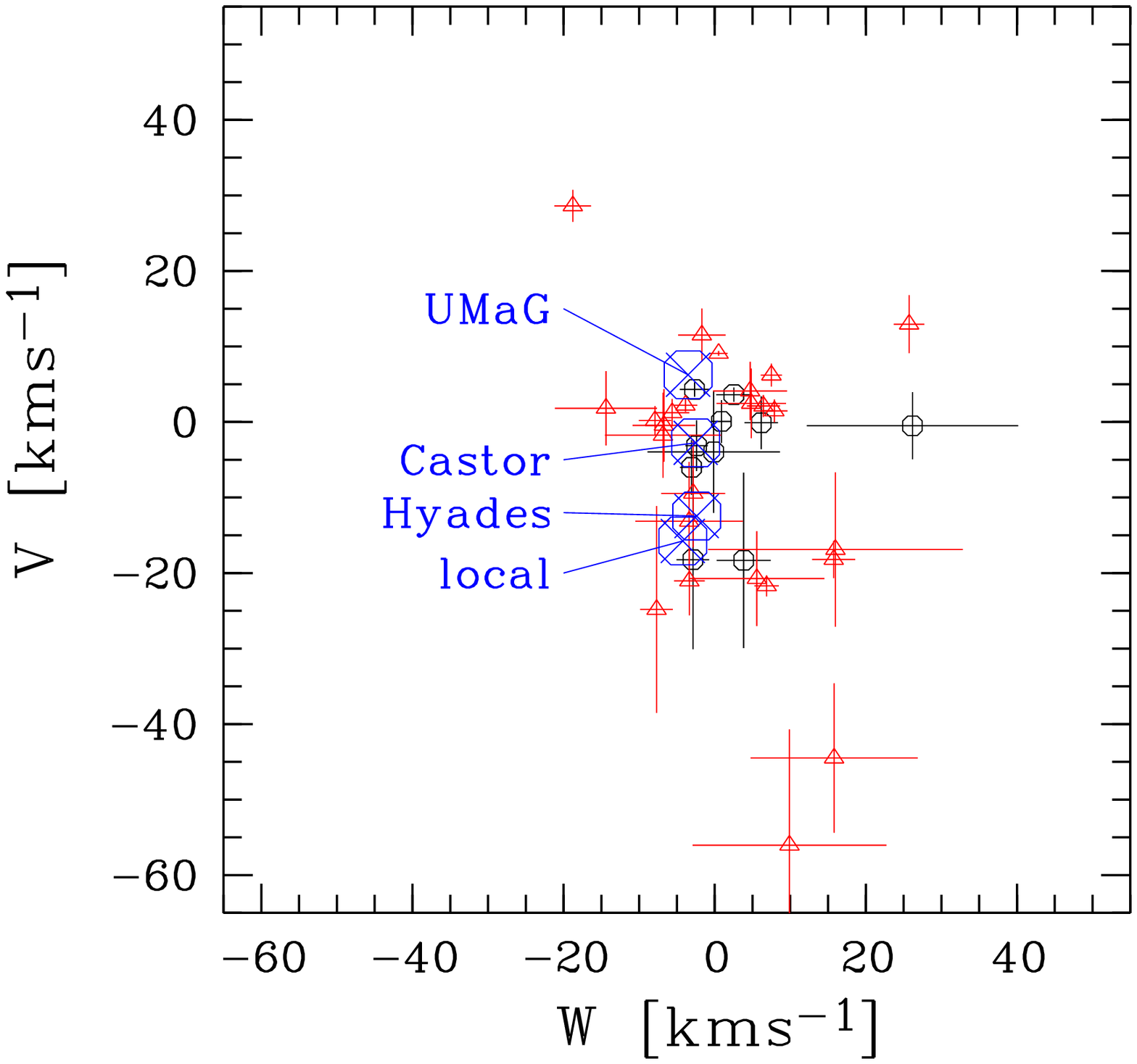,angle=-90,width=8.7cm}
  \end{center}
\caption{Upper Panel: $U-V$ diagram for PMS ($\circ$) and PL\_ZAMS stars
($\bigtriangleup$). Included are the limits for young disk
stars from Eggen (1989) 
and space velocities of the indicated clusters and MGs from Montes et al. (2001).
Lower  panel: $W-V$ diagram for the same sample of stars. Velocities are in the LSR
reference frame.
}
\label{UVeggen}
\end{figure}

For the youngest age groups PMS and 
Pl\_ZAMS an $U-V$-digram is shown in in the upper panel of Fig. \ref{UVeggen}.
The space velocities $U$ and $V$ of these stars cluster with a few exceptions
around two values. The majority  is found close to the velocity of the Castor 
Moving Group (MG).  A second but much smaller fraction is located 
close to the Local Association (``local'' in Fig. \ref{UVeggen}).
The figure also shows that most PMS stars are kinematically consistent 
with the Castor MG but distinct from the Local Association. 

This finding is also consistent with
the velocity component in the direction perpendicular to the galatic plane, $W$, which
is shown in the $W-V$ diagram in the lower panel of Fig. \ref{UVeggen}. 
As for $U$ and $V$ most PMS stars 
are kinematically distinct from the Local Association also in the $W$ space velocity 
component.

\section{Conclusions}
\label{sec:Conclusion}
For the age distribution of the high-galactic latitude coronal sample we found that
about half of the G-K stars are younger than the Hyades.
About 1/3 of the G-K stars is as young or younger than the Pleiades.
A small fraction of less than 10\% of the G-K stars is younger 
than the Pleiades. In contrast to the Pl\_ZAMS group whose members 
are found in all study areas most PMS stars, i.e. 8 out of 10, are located in area~I. 
Only two PMS stars are found in area II and none in the remaining
areas. This could indicate a possible relation of the high $|b|$  PMS stars 
to the Gould Belt indicated in  Fig. \ref{area}. However, the subsample formed 
by combining the stellar age groups PMS and PL\_ZAMS is  
spatially distributed in all directions covered by our study areas. At the same time 
most of its members show similar kinematical parameters independent of 
spatial location. This questions the relation to the Gould Belt. 
Rather, the space velocities are consistent with these stars being members of a loose 
moving group with velocities close to the Castor MG which has an age of 
$\sim200\pm100$\,Myr (\cite{Barrado98}).
If the members of the PMS group of our sample belong indeed to Castor MG this would 
indicate an even larger spread of age in this moving group.



\end{document}